\documentclass[sigconf]{acmart}
\usepackage{amsmath}
\usepackage{graphicx}
\usepackage{amsfonts}
\usepackage{amsthm}
\usepackage{mathtools}
\usepackage{bm}
\usepackage{color}
\usepackage{enumerate}
\usepackage{sgame, tikz}
\usepackage{accents}
\newcommand{\Comments}{0}
\newcommand{\mynote}[2]{\ifnum\Comments=1\textcolor{#1}{#2}\fi}
\newcommand{\yc}[1]{\mynote{red}{[YC: #1]}}
\newcommand{\ignore}[1]{}

\DeclarePairedDelimiter\ceil{\lceil}{\rceil}

\newtheorem{prop}{Proposition}

\theoremstyle{definition}

\acmConference[FAT/ML '17]{Fairness, Accountability, and Transparency in Machine Learning}{August 14--17}{Halifax, Nova Scotia, Canada}
\acmYear{2017}
\copyrightyear{2017x}
\begin{document}
\title[Feedback to Fairness]{Fairness at Equilibrium in the Labor Market}
\author{Lily Hu}
\affiliation{
	\institution{SEAS\\Harvard University}
	\city{Cambridge}
	\state{MA}}
	 \email{lilyhu@g.harvard.edu}
\author{Yiling Chen}
\affiliation{
	\institution{SEAS\\Harvard University}
	\city{Cambridge}
	\state{MA}}
	 \email{yiling@seas.harvard.edu}

\begin{abstract}
Recent literature on computational notions of fairness has been broadly divided into two distinct camps, supporting interventions that address either individual-based or group-based fairness. Rather than privilege a single definition, we seek to resolve both within the particular domain of employment discrimination. To this end, we construct a dual labor market model composed of a Temporary Labor Market, in which firm strategies are constrained to ensure group-level fairness, and a Permanent Labor Market, in which individual worker fairness is guaranteed. We show that such restrictions on hiring practices induces an equilibrium that Pareto-dominates those arising from strategies that employ statistical discrimination or a ``group-blind" criterion. Individual worker reputations produce externalities for collective reputation, generating a feedback loop termed a ``self-fulfilling prophecy." Our model produces its own feedback loop, raising the collective reputation of an initially disadvantaged group via a fairness intervention that need not be permanent. Moreover, we show that, contrary to popular assumption, the asymmetric equilibria resulting from hiring practices that disregard group-fairness may be immovable without targeted intervention. The enduring nature of such equilibria that are both inequitable and Pareto inefficient suggest that fairness interventions are of critical importance in moving the labor market to be more socially just and efficient. 
\end{abstract}
\maketitle

\yc{We don't have departments, just the Paulson School of Engineering and Applied Sciences. I might have used the wrong cls file as the departments don't show up anyway.}

\section{Introduction}

Work in the growing field of algorithmic fairness proposes interventions of discretion on algorithmic decision-makers when issues of bias and discrimination are potentially at stake. The literature is varied but may be broadly categorized as either proposing solutions that defend fairness at the individual level (similar individuals are treated similarly) \citep{dwork2012fairness} or at the group level (groups are awarded proportional representation) \citep{kamishima2011fairness, feldman2015certifying}. This paper constructs a model of discrimination in the labor market along with fairness constraints that address both notions of fairness.




As we focus on the particular domain of labor market dynamics, our paper draws upon an extensive literature in economics. The theory of statistical discrimination, originally set forth in two seminar papers by Phelps \cite{phelps1972statistical} and Arrow \cite{arrow1998has}, explains group-unfair outcomes as the result of rational agent behaviors that lock a system into an unfavorable equilibrium. In the basic model, workers compete for a skilled job with wage $w$. Skill acquisition requires workers to expend an investment cost of $c$, which is distributed according to a function $F$. Thus, a worker's investment decision is an assessment of her expected wage gain compared with her cost of investment. Firms seek information about a worker's hidden \textit{ability} level but can only base their hiring decisions on observable attributes: her noisy \textit{investment} signal and group membership. The firm's response to this informational asymmetry is to update its beliefs about a worker's qualifications by drawing on its prior for her group's ability levels. Therefore, if a firm holds different priors for different groups, it will also set different hiring thresholds. Further, since these distinct thresholds are observed and internalized by workers, they adjust their own investment strategies accordingly---individuals within the unfavored group will lower their investment levels, and individuals in the favored group will continue to invest at a high level. Notably, even when the distribution of costs $F$ is the same for each group\footnote{This has been the standard assumption in the statistical discrimination and labor economics literature since Arrow \cite{arrow1998has}.}, an asymmetric equilibrium can arise in which groups invest at different levels, further informing firms' distinct priors. In other words, rational workers and firms best-respond in ways that exactly confirm the others' beliefs and strategies, and thus, the discriminatory outcome is ``justified."

This equilibria perspective challenges our mission in designing constraints to ensure fairness. For one, given that statistical discrimination and machine learning in general rely on data that harbor historical inequalities, isolated algorithmic interventions that do not consider the dynamics of the particular system in which they are embedded will often fall short of addressing the self-perpetuating nature of biases. We cannot look forward toward a future of fairness without first looking backwards at the conditions that have caused such inequalities. Thus if the observed outcomes themselves are trapped in a feedback loop, fairness constraints should aim to first jolt the system out of its current steady-state, and second, launch it on a path towards a preferable equilibrium. As such, a successful approach must consider fairness \textit{in situ.} This paper presents a dynamic model that is \textit{domain-specific} and a fairness intervention that is \textit{system-wide}.


In our model, workers invest in human capital, enter first a Temporary Labor Market (TLM), and then transition into a Permanent Labor Market (PLM)\footnote{This type of worker movement in a segmented market is common in the labor economics literature. Of these, our work shares a structure and purpose most similar to Kim \& Loury \cite{loury2014collective}. However, notably their model is one of statistical discrimination, while ours explicitly requires group-fairness.}. We use this partition to impose a constraint on firms' hiring practices in the TLM that enforces group-fairness. However, the restriction need not apply in the PLM, and firms naturally select best-response hiring strategies that guarantee individual worker fairness. Our model of the labor market is \textit{reputational}---a worker carries an individual reputation, which is a history of her past job performances, and belongs to one of two groups that has a collective reputation, summarized by the proportion of workers in the group producing ``good" outcomes. 

Working within this model, we show that by imposing an outcomes-based fairness constraint on firms' hiring strategies in the TLM, the resulting steady-state equilibrium in the PLM is group-symmetric and Pareto-dominates the PLM asymmetric equilibria that arise due to either group-blind optimal hiring or statistical discriminatory hiring. Within this model, our fairness intervention is constructed to exploit the complementary nature of individual and collective reputations such that the system produces a feedback loop that incrementally addresses initial inequalities in group social standing. As such, our TLM fairness intervention is not permanent---the guarantee of group-level fairness gradually morphs into a guarantee of individual-level fairness as group equality is restored.

This paper's application of notions of group-based and individual fairness to a reputational model of statistical discrimination in the labor market melds the perspectives and techniques of labor economics with the motivations and framework of algorithmic fairness.\yc{Maybe be explicit about the two literatures.} Our proposed fairness constraints address individual and group fairness in separate treatments in the labor market that are nonetheless linked via a complementarity feedback loop. Thus, these constraints are aimed at creating system-wide conditions of fairness that are self-sustaining. As algorithms are increasingly deployed to make employment-related decisions, we hope our work sheds a light on the equilibrium nature of discriminatory outcomes and may inform the design and implementation of fairness constraints. 

In Section 2, we present our model of labor market dynamics and the imposed fairness constraint. Section 3 contains an overview of the equilibria results of the model under the fairness requirement along with a comparison against equilibria arising from two rational hiring strategies free from such a constraint. The paper ends with a reflection on the general equilibrium tendencies of discrimination and their implications on the design of fairness constraints. We also offer some comments on the dynamic feedback effects that are inherent features of persistent inequalities and the challenges they issue upon future work in algorithmic fairness.

\subsection{Related Work}
Within the algorithmic fairness literature, Zemel et al. \cite{zemel2013learning} address both group and individual notions of fairness by constructing a map of agent data to an intermediate layer of clusters that each preserve statistical parity (ensuring group fairness) while obfuscating protected attributes. A second mapping to classification based on cluster assignments then allows ``similar" agents to be treated similarly. This dual-map approach corresponds to our TLM and PLM fairness constraints. Related work has sought distance metrics to guide the initial mapping \cite{dwork2012fairness}, but since criteria for similarity vary by domain, such general approaches often face obstacles of application. We hope that our paper's concentrated treatment of dynamics in the labor market addresses this concern. We answer a call by Friedler et al. \cite{friedler2016possibility} to specify a particular world view of fairness within a domain and classification task. Our model starts with the important stance of inherent equality between groups. As such, any indications of difference in group ability distribution because of observable investment decisions or outcomes is due to unequal societal standing, producing myriad secondary effects of inequality, rather than fundamental differences in the nature of the individuals. Following the vocabulary used in Friedler et al., our model of group differences in the labor market subscribes to the axiomatic assumption of ``We're all equal." 

\section{Model}
We disentangle the concerns of group-based and individual fairness by constructing a dual labor market that addresses each notion of fairness independently. A firm's hiring process in the Temporary Labor Market (TLM) will guarantee group-based fairness; the mechanisms of the Permanent Labor Market (PLM) will ensure individual fairness. While the labor market is divided as a whole, their dynamics are not separable---workers flow from the TLM to the PLM, wages are labor-market-wide, and individual worker reputations in the PLM produce externalities for the collective group reputations that play a key role in individual's pre-TLM investment decisions. 

\subsection{General Setup}
Consider a society of $n$ workers who pass through the labor market sequentially at times $t = 0, 1, ...$. The labor markets maintain a constant relative size: $m$ proportion of the workers reside in the TLM, and $1-m$ reside in the PLM. Movement is governed by Poisson processes---workers immediately replace departing ones in the TLM, transition from the TLM to the PLM according to the parameter $\kappa$, and leave the PLM at rate $\lambda$.

Each worker belongs to one of two groups $\mu \in \{B, W\}$ within which the distribution of individual abilities is identical, described by the CDF $F(\theta)$. Firms hire and pay workers based on expected performance, awarding wage $w(g_t)$ for a ``good" worker, where $g_t$ gives the proportion of ``good" workers in the PLM at time $t$. This is formalized by assigning workers to either skilled or unskilled tasks with distinct wages. Thus for simplicity, ``bad" workers are also hired but are assigned to the unskilled task and paid a wage normalized to $0$. The wage premium $w(g_t)$ is decreasing in $g_t$, since as the relative supply of ``good" workers increases, imperfect worker substitutability lowers their marginal productivity, thus decreasing wage. We impose a minimum wage $\underaccent{\bar}{w}$ such that $\lim_{g_t \to \infty} w(g_t) = \underaccent{\bar}{w}$ and a maximum wage $\bar{w}$ such that $\lim_{g_t \to 0} w(g_t) = \bar{w}$.

\subsection{Temporary Labor Market}
First, a worker $i$ at time $t$ chooses to invest in human capital $\eta_i > 0$ according to the current wage premium $w(g_t)$\footnote{Workers are boundedly rational and unable to anticipate future wage dynamics.} and her personal cost function for investment, $c_{\pi^{\mu}_{t'}}(\theta_i, \eta_i)$, which is a function of her individual ability $\theta_i$\footnote{``Ability" should be broadly interpreted as encapsulating all personal attributes that bear on success within traditional institutions of education and work.}, group $\mu$ ``reputation," $\pi^{\mu}_{t'}$ where $t'$ represents the time interval $[t-\tau, t]$, and selected level of investment $\eta_i$. The collective reputation gives the proportion of ``good" workers of group $\mu$ in the interval $t'=[t-\tau,t]$, where the parameter $\tau$ controls the time-lag effect of a group's previous generations' reputation on a member's investment cost in the present. $c$ is decreasing in $\theta$, increasing in $\eta$, and $\forall \pi^\mu_{t'} \le \pi^\nu_{t'}, c_{ \pi^\mu_{t'}}(\theta_i, \eta_i)$ is a positive monotonic transformation of $c_{\pi^{\nu}_{t'}}(\theta_i, \eta_i)$. The incorporation of group membership into an individual's cost function is informed by the vast empirical literature that demonstrates the differential externalities produced by groups of differential social standing \citep{bowles2014group}. 

Investment in human capital operates as an imperfect signal, and workers have a hidden true type, qualified or unqualified, $\rho \in \{Q, U\}$ with proportion $\gamma_Q$ being qualified, and $1-\gamma_Q$ being unqualified. For investment level $\eta$, $P(Q|\eta_i) \ge P(Q|\eta_j), \forall \eta_i > \eta_j$. Thus, a firm's \textit{TLM hiring strategy} is a mapping $\mathcal{H}_T: \mathbb{R}$ $\bigtimes$ $\{B,W\} \rightarrow \{0,1\}$ such that the decision for agent $i$ is based only her observed investment level $\eta_i$ and group membership $\mu$. 

\begin{definition}[Group Fairness]
A hiring strategy $\mathcal{H}$ is \textit{group-fair} if and only if for all agents $i$, the event of $i$ being hired is conditionally independent of her group-membership, and thus $$P(W) = P(W | \mathcal{H}(i)) \text{ and } P(B) = P(B | \mathcal{H}(i) )$$
\end{definition}

Group-fair hiring results in employee representation that satisfies statistical parity. We impose this fairness constraint on hiring strategies in the TLM, requiring firms to move beyond ``group-blind" practices. 

\subsection{Permanent Labor Market}
Once hired, worker $i$ exerts on-the-job effort $e \in \{H,L\}$, which stochastically produces an observable outcome $o \in \{G, B\}$ that impacts her individual reputation and thus future reward. $e=L$ is free, but exerting $e=H$ is costly with $c(\theta_i, \rho)$ as a function of qualification status and ability level. \textit{Effort} exertion cost functions here are distinct from the previous \textit{investment} cost functions---the former are pertinent to the PLM and differ by qualification status, whereas the latter relate to the TLM and differ by group membership. Effort is more costly for unqualified individuals and $c(\theta_i,U) \ge c(\theta_i, Q), \forall \theta$, and high effort increases the probability of a good outcome $G$. Thus if $p_{\rho, e}$ gives the probability of achieving outcome $G$ with qualifications $\rho$ and effort level $e$, we have the following inequalities
\begin{align*}
p_{Q,H} > p_{Q,L}; p_{U,H} > p_{U,L}; p_{Q,L} > p_{U,L}
\end{align*}
Notably $p_{Q,H}=p_{U,H}$. Since the effect of qualifications on exerting high effort is already incorporated in its cost, we write both quantities as $p_H$. 

A worker keeps the same TLM job until the Poisson process with parameter $\kappa$ selects her to move into the PLM, where firms are able to observe her history of job outcomes, including her TLM performance. A worker $i$'s time $t$ history is $h^t_i = (o_{i,1}, o_{i,2}, ... , o_{i, t-1})$, a sequence of outcomes that corresponds to her past performances. Firms are boundedly rational and distill a worker's past histories to her ``individual reputation" $\Pi^t_i$, which gives the proportion of outcomes $G \in h^t_i$. A firm's \textit{PLM hiring strategy} is a mapping $\mathcal{H}_P: [0,1] \rightarrow \{0,1\}$ such that the decision regarding agent $i$ is solely a function of $\Pi^t_i$. Similar to the $\mathcal{H}_T$, the optimal $\mathcal{H}_P$ is also based on a threshold strategy such that for a chosen reputation threshold $\hat{\Pi}^t$ at time t, $\forall i$ such that $\Pi^t_i \ge \hat{\Pi}^t$, $\mathcal{H}(i) = 1$, and inversely, $\forall i$ such that $\Pi^t_i < \hat{\Pi}^t$, $\mathcal{H}(i) = 0$.

Levin \cite{levin2009dynamics} also constructs a model of discrimination in which a worker's observed individual reputation is driven by effort level exertions. But we depart from a binary worker's reputation signal and conceive of reputation as a history of previous outcomes. 


\section{Results}
\subsection{Equilibrium Strategies and Steady-States}
Due to feedbacks between individual and collective reputations, multiple equilibria exist. To focus on a particular equilibrium, we suppose that firms in the PLM prefer to only hire workers who consistently exert high effort. We start by describing PLM strategies and then analyze firms' and workers' best responses together.

A worker in the PLM is both history-cognizant as well as future-anticipatory. Agent $i$'s strategy is a selection of time, reputation, wage, and hiring threshold-dependent probabilities of effort exertion $e(\Pi_i^t)$. The discount factor $\delta$ incorporates workers' present-bias as well as the possibility of exiting the market via the $\lambda$-rate Poisson process. Her exertion decisions are chosen according to the expected marginal reward for a $G$ outcome over a $B$ outcome at time $t$.
\begin{align}
R(\theta_i,\Pi^t_i, \hat{\Pi}^t, g_t) &= \mathbb{E}\Big[\sum_{j=t}^ {\infty} (\delta \Phi_j)^j w(g_t) - (e(\Pi_i^j-e(\Pi_i^j-\frac{1}{j-1}))c(\theta_i, \rho)\Big]
 \\
&\text{where } \Phi_j =  \phi \Big({e(\Pi_i^j)}\Big) - \phi \Big({e(\Pi_i^j-\frac{1}{j-1})}\Big) \text{ with } \nonumber
\end{align}
\begin{align}
\phi \Big({e(\Pi_i^t)}\Big) &= \sum_{k=\ceil*{\hat{\Pi}_i^{t+\tau}(t+\tau)-\Pi^t_it}}^{t+\tau} \binom{t+\tau}{k}[e(\Pi_i^k)p_H + (1-e(\Pi_i^k)p_{\rho, L})]^k
\\
&*[e(\Pi_i^k)(1-p_H) + (1-e(\Pi_i^k)(1-p_{\rho, L})]^{t+\tau-k} \nonumber
\end{align}
We simplify $R(\theta_i,\Pi^t_i, \hat{\Pi}^t, g_t)$ to $R_t^\infty$. $\Phi_j$ describes the dynamics of agent $i$'s individual reputation as a function of her effort exertion probability at each possible previous level of reputation, $e(\Pi_i^j)$. The expectation is taken over wage paths $w(g_t)$, and effort exertion is only warranted if $p_HR_t^\infty - c(\theta_i, \rho) \ge p_{\rho, L}R_t^\infty.$

In the PLM, if firms wish to hire all and only workers who consistently exert high effort, their equilibrium strategy is to select a reputation threshold $\hat{\Pi}^t = p_H*t - \Delta(\delta)$ where $\Delta(\delta)>0$ is some small in magnitude function of the discount factor $\delta$. 
Under the particular hiring strategy, there will be a steady-state wage $\tilde{w}$ such that all workers with ability $\theta_\rho > c_\rho^{-1}(\tilde{w})$ will exert effort $H$ even if their history has ``fallen behind" $\hat{\Pi}^t$. Although immediate reward is not guaranteed, in the long-run, exertions of effort will be rewarded.

However, in this setup in which employers observe reputations $\Pi^t_i$ of outcomes up to $t-1$, workers hold a first-mover advantage. A worker $i$ with reputation $\Pi_i^t - \frac{1}{t} > \hat{\Pi}^t$ who has secured a job at time $t$ may find it profitable to exert effort $L$ given that she knows that she will be hired in the next round regardless of her round $t$ outcome. Thus the optimal worker strategy entails exerting effort in a manner that maintains a reputation exactly oscillating around the threshold $\hat{\Pi}^t$. In response, the firm will optimize its hiring threshold $\hat{h}^t = p_H - \Delta(\delta)$ by decreasing $\Delta$ just enough to motivate consistent high effort from these workers. As $\delta \to 1$, $\Delta \to 0$, and $\hat{\Pi}^t = p_H$ such that workers always exert effort if they can afford to do so. All other workers exert low effort in each round. Thus at any time $t$, given the firm's reputation threshold $\hat{\Pi}^t$, its equilibrium PLM hiring strategy $\mathcal{H_P}$ is a mapping such that if and only if a worker $i$ has reputation $\Pi^t_i > \hat{\Pi}^t$, $\mathcal{H_P}(i) = 1$, else $\mathcal{H_P}(i) = 0$.


Suppose $\gamma_Q$ gives the proportion of candidates who are qualified, leaving $1-\gamma_Q$ who are unqualified. Then the proportion of workers in the PLM who produce good outcomes follows the recursive model
\begin{align}
\widehat{g_{t}} = &p_H[1-F(\widehat{\theta_Q})\gamma_Q-F(\widehat{\theta_U})(1-\gamma_Q)]+ p_{Q,L}F(\widehat{\theta_Q})\gamma_Q\label{sseq}
\\
&+p_{U,L}F(\widehat{\theta_U})(1-\gamma_Q) \nonumber
\\
&\text{where } \widehat{\theta_\rho} = c_\rho^{-1}(w({g_{t-1}})(p_H - p_{\rho, L}))
\label{sseq-cost}
\end{align}
To determine wage dynamics, we consult TLM strategies before proceeding to the full equilibrium description.

Since a TLM firm prefers high-ability candidates, optimal hiring follows a threshold strategy: Given a hiring threshold $\hat{\eta}$, $\forall i$ such that $\eta_i \ge \hat{\eta}$, $\mathcal{H_T}(i) = 1$, and inversely, $\forall i$ such that $\eta_i < \hat{\eta}$, $\mathcal{H_T}(i) = 0$. 

However, since firms must abide by the \textit{group fairness} hiring rule, if a firm aims to hire a fraction $\ell$ of all workers, investment thresholds will be implicitly defined and group-specific. 
\begin{prop}
Taken together, firms' equilibrium hiring strategies for the TLM and PLM, $\mathcal{H_T}$ and $\mathcal{H_P}$, both satisfy group fairness.
\end{prop}

Given the time $t$ TLM investment threshold $\hat{\eta}$, all workers $i$ with $c_{\pi^{\mu}_{t'}}(\theta_i, \hat{\eta}) \le w(g_t)$ will invest at exactly the level $\eta_i = \hat{\eta}$ and be successfully hired, while all other workers will invest at level $\eta_i = 0$ and fail to qualify for skilled job positions in the TLM. 

Once in the TLM, workers know that their future PLM opportunities will depend on their observable outcome in the TLM, and as such they exert effort in a one-shot game. Interestingly, the one-shot game produces a TLM equilibrium that is equivalent to its PLM counterpart, and $\overline{g_t}$, the proportion of workers producing good outcomes in the TLM, follows the structure of (\ref{sseq}) and (\ref{sseq-cost}). This is because a single-shot game imposes the same type of pressure as does the stringent threshold history hiring strategy in the PLM. In both, every outcome ``counts." 

Having elaborated upon the dynamics of both the TLM and PLM, we incorporate worker movement and combine the results to obtain a recursive relationship that governs the equilibrium path of workers' performance results, sequence of $\{g_t\}_0^\infty$, from an initial wage $w_0$. Note that the multiplicity of possible firm hiring strategies produces a multiplicity of equilibrium paths, but given that firms are willing to hire only and all workers who consistently exert high effort, firm and worker equilibrium strategies are as previously described, and there exists a unique equilibrium path. 

\begin{theorem} Suppose firms and workers play the following equilibrium strategies: A firm in the TLM hires a proportion $m$ of workers under the fairness constraint. A firm's PLM strategy follows a time-invariant threshold rule $\hat{h} = p_H - \Delta(\delta)$. A worker $i$ of type $\rho$ exerts effort in the TLM or PLM if and only if $c(\theta_i, \rho) \le w(g_{t-1})(p_H-p_{\rho,L})$. 

Under the above conditions, the proportion of workers producing good outcomes at time $t$ follows the recursive relationship
\begin{align}
g_t = p_H[1-F({\theta_Q})\gamma_Q-F({\theta_U})(1-\gamma_Q)]+ p_{Q,L}F({\theta_Q})\gamma_Q
\\
\text{where } {\theta_\rho} = c_\rho^{-1}(w({g_{t-1}})(p_H - p_{\rho, L})) \nonumber
\end{align}
The proportion $g_t$ is equal for each group $B$ and $W$, satisfying group-fairness throughout the labor market.
\end{theorem}


Under the TLM fairness constraint, groups with unequal initial social standing will gradually approach the same reputation level according to time-lag $\tau$. If the TLM fairness intervention occurs at $t=t_1$, and initially $\pi^B_{t_1} < \pi^W_{t_1}$, the function $\pi^B_{t'}$ where $t' = [t-\tau, t]$ is strictly increasing for all $t > t_1$ until some time T when $\pi^B_{T} = \pi^W_{T}$. The ``self-confirming" loop is now co-opted for group $B$'s reputation improvement---collective reputation produces a positive externality, lowering individual group members' cost functions, thus improving investment conditions for future workers, further raising individual and group reputation. 

We next compare this equilibrium under the TLM group-level fairness constraint with equilibria under other rational hiring strategies that are not bound by any notions of fairness and show that under weak conditions, the fairness equilibrium is Pareto-dominant. 

\subsection{Comparative Statics with Unconstrained Hiring Strategies}
Consider a TLM hiring strategy that is individual-based, operating under a pure equal-treatment philosophy. A firm hires a proportion $q$ of workers by setting an investment level $\eta$ defined implicitly as
\begin{align}
q = (1-\sigma_B)(1-F(c_W^{-1}(\eta))+ \sigma_B(1-F(c_B^{-1}(\eta))
\end{align}
where $\sigma_B$ and $1-\sigma_B$ give the proportion of individuals in groups $B$ and $W$ respectively and the function $c_\mu(\theta)$ gives the group $\mu$ investment level function. Writing $c_\mu^{-1}(\eta) = \widetilde{\theta_\mu}$ for $\mu \in \{B,W\}$, these TLM ability thresholds are ranked with respect to the ability threshold under the fairness constraint $\theta$ as $\widetilde{\theta_W} < \theta <  \widetilde{\theta_B}$. 

In the PLM, the steady-state wage $\tilde{w}$ enforces thresholds $c_j^{-1}(\tilde{w}) = \widehat{\theta_\rho}$ for $\rho \in \{Q, U\}$. If $ \widehat{\theta_Q} \in [\theta, \widetilde{\theta_B}]$\footnote{$\widehat{\theta_Q} > \theta$ is not a stringent requirement since it is expected that the PLM threshold is higher than the TLM threshold, else all $Q$ workers would be hired at equilibrium.}, then asymmetric equilibria in which group reputations $\pi^{\mu}$ differ may arise. Firms' TLM hiring strategies inequitably bound the proportion of high-ability workers in group $B$ who are eligible to compete for jobs in the PLM, thus maintaining the reputation gap and differences in group investment costs, producing the ``self-confirming" equilibrium effect.

Further, since $1-F_g( \widehat{\theta_\rho}) < 1-F_f( \widehat{\theta_\rho})$ where $F_g$ and $F_f$ are the ability CDFs under the ``group-blind" and fair regime respectively, firms that demand more workers strictly prefer the equilibria under the fairness constraint. This is because the effective higher ability threshold for group $B$ under this ``group-blind" TLM strategy is inefficient, leaving behind an untapped resource of skilled and qualified individuals who would have otherwise been hired in the PLM. Even the hired workers in group $W$ who are not hired in the fair regime do not fare better, since all such workers have ability level lower than the PLM reputation threshold and are not hired at equilibrium anyway. 

Firms may also select hiring strategies that use statistical discrimination, in which priors regarding a worker's observable attributes (such as group membership) are used to infer a particular individual's hidden attributes. However, when groups face differing underlying investment cost functions, firms that statistically discriminate may push the system toward Pareto-dominated equilibria similar in kind to the asymmetric equilibria in the ``group-blind" case. 

In particular, if TLM firms hold priors $\xi_B$ and $\xi_W$ about the two groups' capabilities, upon observing an agent's group membership $\mu$ and investment level $v$, they will update their beliefs according to: 
\begin{align}
P(S | \mu, v) = \frac{p_S(v) \xi_\mu}{p_S(v) \xi_\mu + (1-\xi_\mu)p_U(v)}
\end{align}
where $p_S(v)$ and $p_U(v)$ gives the probability of a skilled and unskilled worker having investment level $v$ respectively.
 
If $\xi_W > \xi_B$, then $P(S | W, v) > P(S | B, v)$, and the groups face different incentive compatibility constraints. As Coate \& Loury \cite{coate1993will} show, self-confirming asymmetric equilibria also exist under this regime, and lower investment levels within the group with lower social standing are justified by firms' more stringent hiring standards. These TLM choices have ramifications in the PLM that mirror the Pareto-dominated results under ``group-blind" hiring.
 
\section{Discussion}
Given the specific nature of demands in fairness, domain-specific approaches lend themselves to better modeling of the impact an intervention can make on a particular ecosystem. Describing unfair outcomes in employment as caused by rational agent best-response strategies suggests that the field of algorithmic fairness should consider the labor market's inherent dynamic setting in its approach to potential interventions. Fairness constraints that are conceived as isolated procedural checks have a limited capacity to install system-wide fairness that may be self-sustained and long-lasting. The problem of fairness is fundamentally tied to historicity. Within all societal domains in which fairness is an issue, past and current social relations differentially impact subjects, producing distinct sets of resources, options, and opportunities that continue to mark agents' choices and outcomes today. This fact presents a challenge for the standard learning theory formulation of the problem in which agent attributes are treated as \textit{a priori} givens rather than themselves the products of a lineage of previous social choices and conditions. A dynamic model recognizes the powerful ripple effect of the past and calls for a fairness intervention that carries momentum into the future. The labor market as a source of economic opportunity is ripe with positive externalities and is thus an ideal setting for a notion of fairness that is oriented toward a future beyond the short timeline of firm hiring cycles. We argue that fairness conceived in this way is a project that aims to achieve group egalitarianism---an ambition that is not only a worthy goal in itself but one that we show is also socially optimal. 

Our model of individual reputations as a sequence of previous outcomes in the PLM fits within the hiring process regime today, in which employers have increased access to worker data. Since algorithms will be largely responsible for making sense of this individual historical data, future work should consider interventions that are able to sift through a worker's historical data and determine whether a group-based fairness constraint, such as the one considered in this paper in the TLM, should be imposed, or whether a PLM individual-based hiring decision will suffice. Beyond serving efficiency, such practices should protect fairness.


The entry of algorithms into hiring must grapple with a long tradition of explicit and implicit human biases that have rendered the labor market prone to discriminatory practices. We hope that this work can suggest ways that algorithmic fairness interventions can shift hiring strategies towards a better, fairer future.

\bibliographystyle{unsrtnat}
\bibliography{bibliography-fairness}

\begin{thebibliography}{11}
\providecommand{\natexlab}[1]{#1}
\providecommand{\url}[1]{\texttt{#1}}
\expandafter\ifx\csname urlstyle\endcsname\relax
  \providecommand{\doi}[1]{doi: #1}\else
  \providecommand{\doi}{doi: \begingroup \urlstyle{rm}\Url}\fi

\bibitem[Dwork et~al.(2012)Dwork, Hardt, Pitassi, Reingold, and
  Zemel]{dwork2012fairness}
Cynthia Dwork, Moritz Hardt, Toniann Pitassi, Omer Reingold, and Richard Zemel.
\newblock Fairness through awareness.
\newblock In \emph{Proceedings of the 3rd Innovations in Theoretical Computer
  Science Conference}, pages 214--226. ACM, 2012.

\bibitem[Kamishima et~al.(2011)Kamishima, Akaho, and
  Sakuma]{kamishima2011fairness}
Toshihiro Kamishima, Shotaro Akaho, and Jun Sakuma.
\newblock Fairness-aware learning through regularization approach.
\newblock In \emph{Data Mining Workshops (ICDMW), 2011 IEEE 11th International
  Conference on}, pages 643--650. IEEE, 2011.

\bibitem[Feldman et~al.(2015)Feldman, Friedler, Moeller, Scheidegger, and
  Venkatasubramanian]{feldman2015certifying}
Michael Feldman, Sorelle~A Friedler, John Moeller, Carlos Scheidegger, and
  Suresh Venkatasubramanian.
\newblock Certifying and removing disparate impact.
\newblock In \emph{Proceedings of the 21th ACM SIGKDD International Conference
  on Knowledge Discovery and Data Mining}, pages 259--268. ACM, 2015.

\bibitem[Phelps(1972)]{phelps1972statistical}
Edmund~S Phelps.
\newblock The statistical theory of racism and sexism.
\newblock \emph{The american economic review}, 62\penalty0 (4):\penalty0
  659--661, 1972.

\bibitem[Arrow(1998)]{arrow1998has}
Kenneth~J Arrow.
\newblock What has economics to say about racial discrimination?
\newblock \emph{The journal of economic perspectives}, 12\penalty0
  (2):\penalty0 91--100, 1998.

\bibitem[Loury and Kim(2014)]{loury2014collective}
Glenn~C Loury and Young-Chul Kim.
\newblock Collective reputation and the dynamics of statistical discrimination.
\newblock 2014.

\bibitem[Zemel et~al.(2013)Zemel, Wu, Swersky, Pitassi, and
  Dwork]{zemel2013learning}
Rich Zemel, Yu~Wu, Kevin Swersky, Toni Pitassi, and Cynthia Dwork.
\newblock Learning fair representations.
\newblock In \emph{Proceedings of the 30th International Conference on Machine
  Learning (ICML-13)}, pages 325--333, 2013.

\bibitem[Friedler et~al.(2016)Friedler, Scheidegger, and
  Venkatasubramanian]{friedler2016possibility}
Sorelle~A Friedler, Carlos Scheidegger, and Suresh Venkatasubramanian.
\newblock On the (im) possibility of fairness.
\newblock \emph{arXiv preprint arXiv:1609.07236}, 2016.

\bibitem[Bowles et~al.(2014)Bowles, Loury, and Sethi]{bowles2014group}
Samuel Bowles, Glenn~C Loury, and Rajiv Sethi.
\newblock Group inequality.
\newblock \emph{Journal of the European Economic Association}, 12\penalty0
  (1):\penalty0 129--152, 2014.

\bibitem[Levin et~al.(2009)]{levin2009dynamics}
Jonathan Levin et~al.
\newblock The dynamics of collective reputation.
\newblock \emph{The BE Journal of Theoretical Economics}, 9\penalty0
  (1):\penalty0 1--25, 2009.

\bibitem[Coate and Loury(1993)]{coate1993will}
Stephen Coate and Glenn~C Loury.
\newblock Will affirmative-action policies eliminate negative stereotypes?
\newblock \emph{The American Economic Review}, pages 1220--1240, 1993.

\end{thebibliography}

\end{document}